\documentclass[12pt]{article}
\usepackage{geometry}
\usepackage{epsfig}
\usepackage{color}
\usepackage{graphicx}
\usepackage{amsmath}
\usepackage{amsfonts}
\usepackage{verbatim}
\usepackage{relsize}
\usepackage{bm,bbm}
\usepackage[normalem]{ulem}
\usepackage{array}
\usepackage{booktabs}
\usepackage{kbordermatrix}
\usepackage{multirow,multicol}
\usepackage{rotating}
\usepackage{url}
\usepackage{float}
\usepackage{amsthm}
\usepackage{subfig}
\usepackage{qtree}
\usepackage[edges]{forest}
\usepackage{amsthm}
\usepackage{bigints}
\usepackage{xcolor,colortbl}
\usepackage{hyperref}
\usepackage{caption}
\usepackage{dsfont}
\usepackage{enumitem}
\usepackage[]{algorithmic}
\usepackage[0]{algorithm}

\floatname{algorithm}{}
\definecolor{gris}{gray}{0.5}

\newcommand{\ts}{\textsuperscript}

\newcolumntype{C}[1]{>{\centering\let\newline\\\arraybackslash\hspace{0pt}}m{#1}}
\newcolumntype{L}[1]{>{\raggedright\let\newline\\\arraybackslash\hspace{0pt}}m{#1}}
\newcolumntype{R}[1]{>{\raggedleft\let\newline\\\arraybackslash\hspace{0pt}}m{#1}}

\usepackage{appendix,apptools,titlesec}

\usepackage[numbers]{natbib}
\def\boxit#1{\vbox{\hrule\hbox{\vrule\kern6pt
          \vbox{\kern6pt#1\kern6pt}\kern6pt\vrule}\hrule}}

\definecolor{orange}{rgb}{1,0.5,0}
\definecolor{MyDarkBlue}{rgb}{0,0.08,0.45}

\usepackage{eurosym}

\begin{document}

\title{Modeling and measuring incurred claims risk liabilities for a multi-line property and casualty insurer\thanks{
 Financial support from NSERC (Godin: RGPIN-2017-06837, Mailhot: RGPIN-2015-05447) and MITACS (Araiza Iturria, Godin and Mailhot: IT12099) is gratefully acknowledged.}}

\author{Carlos Andr\'es Araiza Iturria\footnote{
University of Waterloo, Department of Statistics and Actuarial Science,
200 University Ave W, Waterloo, Ontario, Canada, N2L 3G1,
\textit{caraizai@uwaterloo.ca}}, 
Fr\'ed\'eric Godin\footnote{
Concordia University, Department of Mathematics and Statistics, 
1455 Boulevard de Maisonneuve O, Montr\'eal, Qu\'ebec, Canada, H3G 1M8,
\textit{frederic.godin@concordia.ca}},
M\'elina Mailhot\footnote{
Concordia University, Department of Mathematics and Statistics, 
1455 Boulevard de Maisonneuve O, Montr\'eal, Qu\'ebec, Canada, H3G 1M8,
\textit{melina.mailhot@concordia.ca}}
}
\maketitle


\begin{abstract}

We propose a stochastic model allowing property and casualty insurers with multiple business lines to measure their liabilities for incurred claims risk and calculate associated capital requirements. Our model includes many desirable features which enable reproducing empirical properties of loss ratio dynamics. For instance, our model integrates a double generalized linear model relying on accident semester and development lag effects to represent both the mean and dispersion of loss ratio distributions, an autocorrelation structure between loss ratios of the various development lags, and a hierarchical copula model driving the dependence across the various business lines. The model allows for a joint simulation of loss triangles and the quantification of the overall portfolio risk through risk measures. Consequently, a diversification benefit associated to the economic capital requirements can be measured, in accordance with IFRS 17 standards which allow for the recognition of such benefit. The allocation of capital across business lines based on the Euler allocation principle is then illustrated. The implementation of our model is performed by estimating its parameters based on a car insurance data obtained from the General Insurance Statistical Agency (GISA), and by conducting numerical simulations whose results are then presented.
\end{abstract}

\medskip\noindent 
KEYWORDS: IFRS 17,  Loss triangles, Double Generalized Linear Models,  Hierarchical copulas, Risk measures, Capital allocation.


\newpage
\baselineskip 1.5em

\section{Introduction}
An important task in the practice of property and casualty insurance is the prediction of future claims arising from incurred liabilities. These claims are commonly known as the unpaid claim liabilities. Such predictions are required for multiple purposes such as reserves calculations, financial reporting and the determination of economic capital requirements. The objective of this article is to provide a model allowing a multi-line insurance company to forecast its unpaid claim liabilities, taking into account possible dependencies between business lines. 
The model combines numerous desirable features and reproduces empirical characteristics of loss ratio dynamics. 

First, a Tweedie distributed Double Generalized Linear Model (DGLM) is used to represent the marginal distribution of loss ratios for each business line. The proposed model is flexible and allows for the fluctuation of both the mean and dispersion across accident semesters and development lags. Furthermore, the Tweedie distribution allows for a mass at zero, representing the situation where no loss is observed, which is frequent in property and casualty insurance. The Tweedie family of distributions was introduced in \citet{Tweedie1984}. The flexibility of the Tweedie family and its ability to model null losses have made it an attractive distribution for loss reserving as in \citet{Avanzi2016} and \citet{Smolarova2017}.

Generalized linear models (GLM) are commonly used in the insurance industry to forecast future claims due to their advantageous trade-off in terms of flexibility, parsimony and ease of interpretation. 
A generalization of the GLM model called \textit{double generalized linear model} (DGLM) is considered in \citet{Smythjorgensen2002} as an alternative when the number of claims is not available. In recent actuarial literature, applications in non-life insurance of Tweedie DGLM can be found in \citet{Boucher2011} and \citet{Andersen2017}.
The latter generalization enables modeling the variability of the dispersion parameter jointly with the mean, instead of considering a fixed dispersion parameter as in traditional GLMs.
The current work makes the assumption that loss ratios for a given accident semester and business line are autocorrelated across development lags. This assumption has been explored in \citet{Sarka2013} where different correlation structures are compared in a claim reserving setting. An application of GLMs with correlated observations in the context of property and casualty insurance can be found in \citet{Smolarova2017} which illustrates the use of such models for insurance pricing. 
Due to the autocorrelation assumption of our model, the estimation procedure used in the current paper relies on Generalized Estimating Equations which are presented among others in \citet{LongitudinalData} or \citet{Hardin2013}, and more specifically in an insurance setting in \citet{Smolarova2017}. 

Another notable characteristic of the model developed herein is its convenient specification of the dependence between losses of the various business lines. Indeed, a hierarchical copula is embedded in the model, which allows for a flexible and easily interpretable representation of the global dependence structure. Copulas have been recently gaining in popularity due to recent developments in the copulas theory and to increases in computational power provided by modern computers. In \citet{Shi2011}, a Gaussian copula is used to measure dependence between personal and commercial auto lines. A hierarchical copula model is used in \citet{Burgi2008} to represent the dependence in an insurance portfolio and to study the calculation of the diversification benefit for the insurer. A rank-based hierarchical copula method dealing with multiple property and casualty insurance lines with different parametric copula families is used and compared to a nested Archimedean copula in \citet{Cote2016}. To perform simulations out of the hierarchical copula model, \citet{Arbenz2012} establishes rigorous mathematical foundations and adapts the Iman-Conover reordering algorithm used in the current paper.

The usefulness of the model presented here is highlighted in the context of the new IFRS 17 accounting standards. Indeed, such standards require the calculation of a quantity referred to as the \textit{risk adjustment for non-financial risks}, which will be detailed subsequently. Among the admissible methods for the calculation of the latter quantity, some require a specification of the entire portfolio loss distribution. Our modeling framework allows for the construction of such a distribution; it even provides the joint distribution of unpaid claim losses over all business lines. Hence, the current work illustrates how our loss triangles prediction model can be leveraged within a stochastic simulation to estimate the latter joint distribution, and therefore obtain estimates for the risk adjustment for non-financial risk and capital requirements. The allocation of reserves and capital requirements across the business lines based on the Euler allocation principle is also illustrated.

The paper is organized as follows. Section \ref{section:IFRS} discusses the valuation of insurance liabilities under the new IFRS 17 standards. Section \ref{section:Data} describes the Canadian automobile insurance dataset used in the current study. In Section \ref{section:Model}, the prediction model for loss ratios of a multi-line property and casualty insurer is presented. 
In Section  \ref{section:Imple}, the estimation and simulation of the model are discussed.
In Section \ref{section:Numerical}, a numerical application of the model is illustrated.  A stochastic simulation involving the use of traditional risk measures is compared to a cost of capital approach with respect to the calculation of capital requirements, including its allocation across the various business lines. Section \ref{section:Conclusions} concludes.


\section{IFRS 17 standards in insurance}
\label{section:IFRS}
The International Accounting Standards Board (IASB), an independent international non-profit group of experts in accounting and financial reporting, issued in May 2017 the IFRS 17 \textit{Insurance Contracts} standards, a new set of accounting standards for insurance contracts superseding the current regulatory framework IFRS 4. IFRS 17 \textit{Insurance Contracts} establishes principles for the recognition, measurement, presentation and disclosure of insurance contracts.

The effective date of IFRS 17 has officially been set by the IASB to January 1\ts{st}, 2023,\footnote{The IASB had originally proposed the implementation of IFRS 17 to be effective in 2021, but it has been delayed following public consultation.}
meaning March 31\ts{st}, 2023 would correspond to the first quarter of reporting under IFRS 17.

The unpaid claim liabilities, which are a very important part of the liabilities found in the balance sheet of a property and casualty insurer, are referred to under IFRS 17 as the Liabilities for Incurred Claims (LIC). The LIC represent insurance events that have already occurred, but for which the claims have not been reported or have not been fully settled. A paramount duty for insurers in the upcoming years will consist in measuring the LIC in a manner that is consistent with IFRS 17 \textit{Insurance Contracts} standards. LIC are measured with the General Model which establishes in paragraph 32 of \citet{IFRS17} that upon initial recognition, a group of insurance contracts should be measured as the sum of:

\begin{itemize}
	\item The fulfillment cash flow (FCF), which includes:
	\begin{itemize}
		\item Estimates of future cash flows,
		\item An adjustment to reflect the time value of money and the financial risks related to the future cash flows,
		\item A risk adjustment for non-financial risk.
	\end{itemize}
	\item The contractual service margin (CSM).
\end{itemize}

The CSM represents the unearned profit that the insurer will recognize as it provides services in the future, as stated in paragraph 38 of \citet{IFRS17}. CSM applies for unexpired coverage. It is excluded from the scope of this work. 

The risk adjustment for non-financial risks is the compensation the entity requires for bearing the uncertainty of the amount and timing of the cash flows arising from non-financial risks associated to claim losses, see Paragraph B88 of \citet{IFRS17}. The choice of the methodology to calculate the risk adjustment for non-financial risks is not prescribed by IFRS 17 standards. A few possible approaches, either based on the Cost of Capital (CoC) or risk measures (e.g. VaR and TVaR) are considered in subsequent sections of the current study.


\section{Data}
\label{section:Data}
The dataset used in our analysis comes from the General Insurance Statistical Agency (GISA) and corresponds to data for the entire Canadian automobile industry. The dataset contains entries for two different property and casualty insurance lines, namely personal auto (PA) and commercial auto (CA), and for three regions: Ontario (ON), Alberta (AB) and Atlantic Canada (ATL)\footnote{Atlantic Canada is made up of four provinces: Prince Edward Island, New Brunswick, Nova Scotia and Newfoundland \& Labrador.}. Incremental incurred claim amounts and earned premiums are provided in the dataset.

Loss and Expense (L\&E) semestrial claim amounts are available for each insurance line and region combination from the first semester of 1997 to the second semester of 2017. Data from before 2003 are discarded, since notes on historical claims are only available starting from the first semester of 2003. Therefore, fifteen years of information are taken into account, i.e. data from 2003 to 2017.

In order to work with stationary data, incremental semestrial loss amounts are scaled by the premium for the associated accident semester,
which provides loss ratios. Indeed, the use of loss ratios instead of loss amounts removes the need to quantify trends related to year-to-year changes in exposure.

Previously observed loss ratios can be presented graphically in an upper triangle array, commonly known as a run-off triangle. This presentation can be done in two different ways; using either cumulative claims 
or incremental claims. 
Incremental claims are used in the current study.  
Indeed, for a business line $k$, an accident semester $i$, and a development lag $j$,\footnote{The development lag is the number of semesters between the occurrence of an accident and the date in which the final payment is made (closure of case).} the loss ratio $Y^{(k)}_{i,j}$ is defined as
\begin{equation*}
Y^{(k)}_{i,j}=\cfrac{C^{(k)}_{i,j}-C^{(k)}_{i,j-1}}{p_i^{(k)}},\quad C_{i,0}=0,
\end{equation*}
where $p_i^{(k)}$ is the amount of premiums collected for accident semester $i$ and business line $k$, and $C^{(k)}_{i,j}$ represents the cumulative claims associated with accident semester $i$ obtained until development lag $j$ for business line $k$.
The loss ratios run-off triangles are then used in the subsequent modeling steps.

For illustrative purposes, the current study considers a fictitious insurer whose exposure is the content of the entire aforementioned GISA dataset.




\section{Model}
\label{section:Model}

Consider an insurance portfolio composed of $K$ possibly dependent business lines.
The main objective is to model the joint distribution of loss ratios $Y^{(k)}_{i,j}$ for all possible values of accident semester $i$, development lag $j$ and business line $k$. Then, one can predict loss ratios for future periods based on the observed ones. We obtain the run-off triangle of all semestrial loss ratios observed in a period of fifteen years, i.e. for all accident semesters $i=\{1,2,\ldots,I\}$ with $I=30$, and development lag $j=\{1,2,\ldots,J\}$ with $J=30$ such that $i+j \leq I+1$. The upper and lower triangles are defined as
\begin{eqnarray*}
\mathcal{T}_U &:=& \{ (i,j) : i\in \{1,2,\ldots,I\}, j\in \{1,2,\ldots,J\} ,i+j \leq I+1 \},
\\ \mathcal{T}_L &:=& \{ (i,j) : i\in \{1,2,\ldots,I\}, j\in \{1,2,\ldots,J\} ,i+j > I+1 \},
\end{eqnarray*}
which represent respectively loss ratios that are observable, and those which need to be predicted to perform loss reserving.
Once a model is fitted to loss ratios in $\mathcal{T}_U$, loss ratios from $\mathcal{T}_L$ can be forecasted. The latter procedure is known as completing the square.

In Section~\ref{section:Marginal}, the marginal distribution of loss ratios $Y^{(k)}_{i,j}$ is modeled for each $k$, $i$ and $j$. 
Then, the dependence between loss ratios across development lags and business lines is modeled in subsequent sections. 

\subsection{Marginal distribution model}
\label{section:Marginal}

For the marginal distribution of loss ratios, a Tweedie distributed DGLM is considered. A DGLM model is a generalization of a GLM where both the mean and the dispersion parameters are dependent on explanatory variables. Explaining the dispersion parameter with a nested GLM adds flexilibity to the model. The Tweedie DGLM is equivalent to the combination of a GLM for modeling the frequency parameter and of another GLM to quantify severity in the classic actuarial Compound Poisson-Gamma model under certain conditions described in \citet{Quijano2011}. As explained in \citet{Smythjorgensen2002}, an additional benefit of using a DGLM with a Tweedie distribution is that it allows handling the case where the claim count has not been observed or recorded, or is not reliable; Tweedie distributions are typically mixed distributions with a positive mass at zero. 

Before introducing the formal marginal model, we introduce the Tweedie distribution. A variable $Y$ is said to have the Tweedie distribution $TW_{p}(\mu,\phi)$ if its density is given by

\begin{equation*}
	f_Y(y;\mu,\phi,p)= a(y;\phi,p) \exp\left[\frac{1}{\phi}\left(y\cfrac{\mu^{1-p}}{1-p}-\cfrac{\mu^{2-p}}{2-p}\right)\right], \quad y>0
\end{equation*}
where $$a(y;\phi,p)=\mathlarger{\sum}_{r=1}^{\infty}\left[\cfrac{\phi^{p-1}y^{\ell}}{(2-p)(p-1)^{\ell}}\right]^r\cfrac{1}{r!\Gamma(r \ell)y},$$
with $\ell=-\frac{2-p}{1-p}$ and $\Gamma(z)=\int^\infty_{0} x^{z-1} e^{-x} dx$ being the Gamma function. The point of mass at zero has a probability provided by
	\begin{equation*}
	f_Y(0;\mu,\phi,p)=\exp\left[-\cfrac{\mu^{2-p}}{\phi(2-p)}\right].
	\end{equation*} 
The expectation and variance of $Y$ are respectively $\mathbb{E}[Y]=\mu$ and $\text{Var}[Y]=\phi \mu^p$.



\begin{figure}[!tbp]
  \centering
  \subfloat[Mean parameters.]{\includegraphics[scale=0.215]{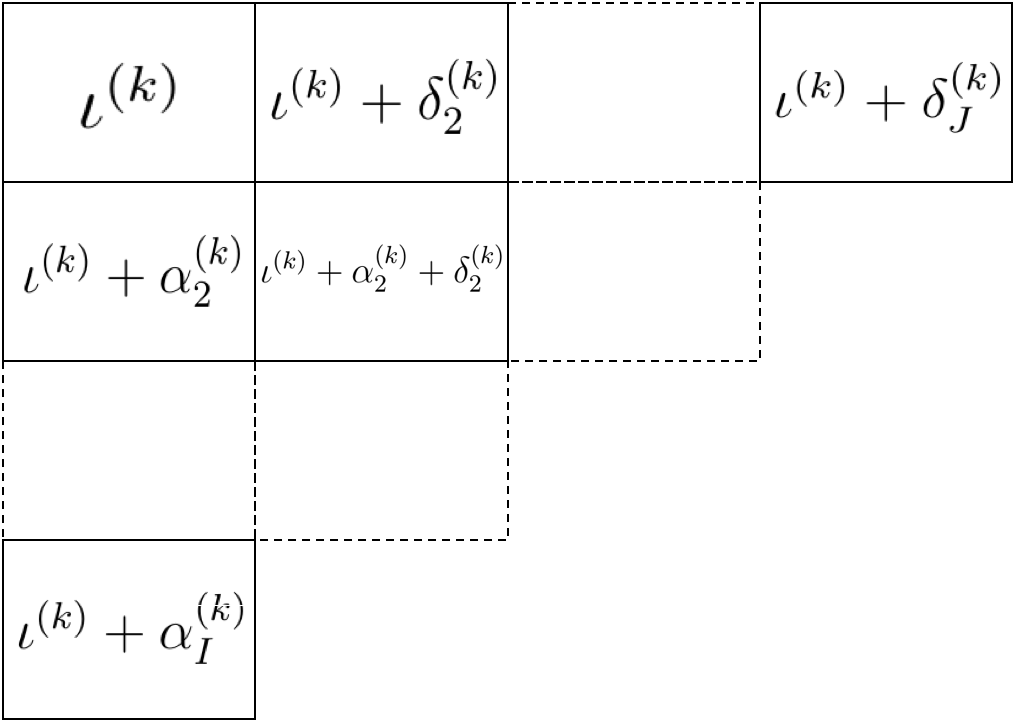}\label{fig:glmeqn}}
  \hfill
  \subfloat[Dispersion parameters.]{\includegraphics[scale=0.215]{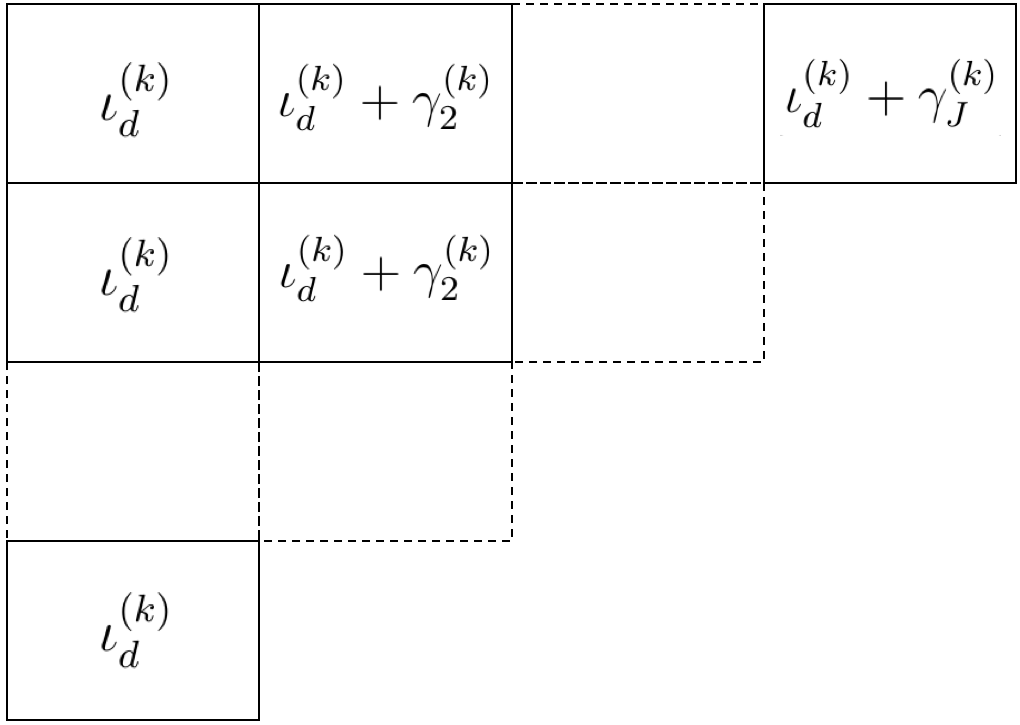}\label{fig:dglm}}
  \caption{Loss triangle's coefficients for the mean GLM model (left panel) and the dispersion GLM model (right panel). Each row of the loss triangle corresponds to an accident semester, whereas each column corresponds to a development lag.}
\end{figure}

In our model, the set of predictors contains exclusively deterministic dummy variables indicating the current accident semester $i$ and development lag $j$. The assumption for our model is that
$Y^{(k)}_{i,j} \sim TW_{p^{(k)}}(\mu_{i,j}^{(k)},\phi_{j}^{(k)})$
with
\begin{eqnarray}
g(\mu^{(k)}_{i,j}) &=&\iota^{(k)}+\alpha^{(k)}_{i}+\delta^{(k)}_{j}, \label{eqn:meanlink}
\\ g(\phi^{(k)}_{j})&=& \iota_d^{(k)}+\gamma^{(k)}_{j}, \label{eqn:dispmodel}
\end{eqnarray}
where constants $\iota^{(k)}$ and $\iota_d^{(k)}$ are respectively the intercept for the mean and dispersion equations, the constants $\alpha^{(k)}_{i}$ and $\delta^{(k)}_{j}$ represent respectively the accident semester and development lag effects for the mean equation, and the constants $\gamma^{(k)}_{j}$ reflect the development lag effect for the dispersion parameter. For any given business line, the dispersion parameter is therefore assumed to depend only on the development lag, and not on the accident semester. This decision is taken to avoid an over-parametrization of the model. Furthermore, unreported verification performed by the authors indicated that including an accident semester effect in the dispersion parameter has a limited impact. Figures \ref{fig:glmeqn} and \ref{fig:dglm} illustrate parameters driving respectively the mean and dispersion of each entry of the loss triangle.

The log-link function $g(x)=\text{log}(x)$ is used for both the mean and dispersion equations, which is a standard choice.

In actuarial literature as in \citet{Shi2011, Cote2016, Smolarova2017} insurance data is usually available for 10 years, making up a total of 55 loss ratios in their data set. Moreover, the GLM they consider has 20 parameters, i.e. the ratio of the number of parameters over the number of data points is 0.36.
Within the 15 years of loss ratio data from the current study, we are dealing with 465 data points for each line of business. Moreover, the DGLM we consider has 89 parameters per business line. Thus, the ratio of the number of parameters over the number of data points is 0.19. Our model can therefore be considered relatively parsimonious in comparison to literature benchmarks; it should therefore not be more prone to overfitting that the latter. Nevertheless, the number of parameters versus number of observations ratio is still considerably high in comparison to many other applications in statistics; care must be applied during the review of the calibration in practice to ensure the variability of loss ratios is not under-estimated due to over-fitting, which would be inconvenient from a risk quantification standpoint.

The marginal distribution model for each business line $k$ involves the following parameters to be estimated: the Tweedie index $p^{(k)}$, intercepts $\iota^{(k)}$ and $\iota_d^{(k)}$, accident semester effects $\alpha^{(k)}_{i}, i=1,\ldots,I$, and development lag effects $\delta^{(k)}_{j}$ and $\gamma^{(k)}_{j}$, $j=1,\ldots,J$. 


\subsection{Dependence within a business line}
\label{section:DepWithin}


The next step in the model construction consists in specifying the dependence structure of loss ratios over the various development lags within a given business line for a given accident semester. The objective is to remove the marginal effects, which allows analyzing the dependence across business lines in further steps.

The first assumption made within the model is that loss ratios from different accident semesters are independent. Thus, we assume that for any business lines $k_1,k_2$ and development lags $j_1,j_2$, when two different accident semesters $i_1 \neq i_2$ are considered, the associated loss ratios $Y^{(k_1)}_{i_1,j_1}$ and $Y^{(k_2)}_{i_2,j_2}$ are independent. This is a standard assumption in the literature, see for instance \citet{Avanzi2016} or \citet{Cote2016}.

Within any given accident semester $i$, for a fixed business line $k$, a dependence structure across development lags is considered. The assumption is that the correlation between loss ratios of development lags $j$ and $j'$ is given by $\text{cor}(Y^{(k)}_{i,j},Y^{(k)}_{i,j'})=\rho^{|j-j'|}_k$ for some constants $\rho_1,\ldots,\rho_K$; we assume that the correlation of loss ratios decreases exponentially as the distance between their respective development period increases. The correlation intensity differs for each line of business, but remains constant over the different accident semesters. The Pearson correlation only measures linear dependence. As such, possibly multiple dependence structures (e.g. copulas) could lead to such correlation structures over the development lags. We choose not to explicitly specify the dependence structure of the loss ratios across the development lag dimension further than through its correlation structure. Note that since predictors in the DGLM model from the current paper are deterministic dummy variables, the dependence structure of loss ratios $Y^{(k)}_{i,j}$ carries over to scaled innovations defined in the next section.


\subsection{Dependence between business lines}
\label{section:Dependence}

The remaining part of the model specification consists in detailing the dependence structure of loss ratios between business lines. A convenient approach to represent this dependence structure in an easily interpretable way consists in using hierarchical copula models (HCM). Such a dependence structure is assumed to hold on the decorrelated loss ratio innovations on which autocorrelation impacts were removed.

For that purpose, define the scaled innovations $\tilde{Y}^{(k)}_{i,j}$ for loss ratios through
\begin{equation*}
    \tilde{Y}^{(k)}_{i,j} = \frac{ Y^{(k)}_{i,j} - \mathbb{E} \left[Y^{(k)}_{i,j} \right]}{ \sqrt{\text{Var} \left[Y^{(k)}_{i,j} \right]}}.
    \label{theo:scaledinnov}
\end{equation*}
\citet{Jorgensen1997} shows the following result allowing to approximate the distribution of scaled innovations by a normal distribution when $Y^{(k)}_{i,j}$ is Tweedie distributed:
\begin{equation}
\label{theo:convergnorm}
\tilde{Y}^{(k)}_{i,j} \xrightarrow{d} N(0,1)  \quad \text{as} \quad \phi\rightarrow 0,
\end{equation}
where $\xrightarrow{d}$ denotes convergence in distribution. Define the column vector $\tilde{\mathbf{Y}}^{(k)}_{i} = \left[\tilde{Y}^{(k)}_{i,1}, \ldots, \tilde{Y}^{(k)}_{i,J}\right]^\top$ which contains loss ratio scaled innovations for all development lags associated with business line $k$ and accident semester $i$. As explained in Section \ref{section:DepWithin}, the correlation matrix of $\tilde{\mathbf{Y}}^{(k)}_{i}$ is given by $R_{k,J}$ which is defined as
\begin{equation}
\label{eqn:corrmatrixfull}
R_{k,J}=
\begin{bmatrix}
1 & \rho_k & \rho^2_k & \ldots&\rho^{J-1}_k  \\
\rho_k & 1 & \rho_k&\ldots&\rho^{J-2}_k  \\
 \rho^2_k & \rho_k & 1&\ldots&\rho^{J-3}_k\\
 \vdots & \vdots & \vdots &\ddots&\vdots\\
 \rho^{J-1}_k & \rho^{J-2}_k & \rho^{J-3}_k&\ldots&1\\
\end{bmatrix}.
\end{equation}

Define $L^{-1}_k$ being the inverse of the lower triangle matrix in the Choleski decomposition of $R_{k,J}$, and $\mathbf{U}^{(k)}_{i} = L^{-1}_k \tilde{\mathbf{Y}}^{(k)}_{i}$ being the decorrelated innovations vector for accident semester $i$ and business line $k$. Elements of the vector $\mathbf{U}^{(k)}_{i}$, denoted respectively $U^{(k)}_{i,1},\ldots,U^{(k)}_{i,J}$, are uncorrelated.

The main assumption about the dependence between business lines is that for any accident semester $i$ and development lag $j$, the copula representing the dependence between decorrelated innovations $U^{(1)}_{i,j},\ldots,U^{(K)}_{i,j}$ does not depend on $i$ nor $j$. The model selected to characterize such dependence is presented in the next section.


\subsection{The hierarchical copula for dependence between business lines}
\label{section:Hierarchical}

Hierarchical copulas are models which involve sequentially specifying the dependence between subgroups of the population and eventually obtain a dependence model between all subgroups.
They are convenient dependence models because they are easy to estimate, validate and interpret. 
Moreover, such models are adapted to frameworks where there exists a natural order in which subgroups can be aggregated.
Such an approach is appropriate in an insurance setting where the portfolio is already subdivided, for instance by geographical regions, dependence on legislation \citet{Burgi2008}, similarity of insurable risk types \citet{Shi2011}, or according to some dependence distance metric as in \citet{Cote2016}.



In order to ease the interpretation of the dependence structure, a bivariate approach is chosen. The six lines of business from the GISA dataset are paired first through a geographical criterion; for each geographical region, the personal and commercial auto lines are linked together through \textit{first level copulas}. The first level copulas $C_1, C_2$ and $C_3$ represent the dependence between the personal and commercial auto decorrelated innovations 
respectively for Ontario, Alberta and Atlantic Canada. 
Then, for the three regional groups obtained, the sum of decorrelated innovations associated with each cluster is considered:
\begin{equation*}
\mathcal{U}^{(\text{ON})}_{i,j} = U^{(1)}_{i,j} + U^{(2)}_{i,j}, \quad 
\mathcal{U}^{(\text{AB})}_{i,j} = U^{(3)}_{i,j} + U^{(4)}_{i,j}, \quad 
\mathcal{U}^{(\text{ATL})}_{i,j} = U^{(5)}_{i,j} + U^{(6)}_{i,j},
\end{equation*}
where for each cluster the first and second components represent respectively the personal and commerical auto lines.

\begin{figure}[H]
	\centering
\captionsetup{justification=centering}
	\begin{forest}
	forked edges,
	for tree={
		parent anchor=south,
		child anchor=north,
		if n children=0{
			tier=terminal,
		}{},
	}
	[$C_5$ [,inner sep = -0[$C_1$ [$U^{(1)}_{i,j}$] [$U^{(2)}_{i,j}$] ]]
	[$C_4$ [$C_2$ [$U^{(3)}_{i,j}$] [$U^{(4)}_{i,j}$ ] ]
	[$C_3$ [$U^{(5)}_{i,j}$] [$U^{(6)}_{i,j}$ ]] ] ]
\end{forest}
\caption{Hierarchical copula model used in the current model}
\label{fig:imanconover}
\end{figure}
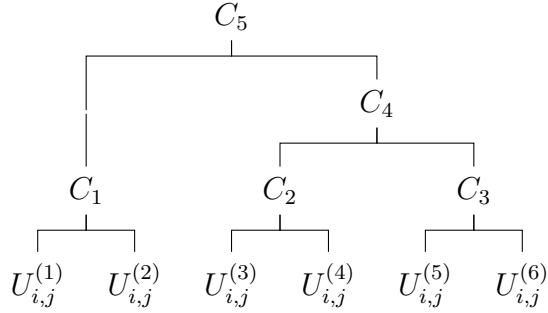

The subsequent level pairing criteria are determined as in \citet{Cote2016} by clustering the most dependent regions based on each pair's Kendall $\tau$. More precisely, a second level copula $C_4$ is incorporated to link summed decorrelated innovations from the Alberta and Atlantic clusters, i.e. the copula $C_4$ represents the dependence model between $\mathcal{U}^{(\text{AB})}_{i,j}$ and $\mathcal{U}^{(\text{ATL})}_{i,j}$. Finally, summing the decorrelated innovations within the Alberta-Atlantic cluster through
\begin{equation*}
\mathcal{U}^{(\text{AB+ATL})}_{i,j} = \mathcal{U}^{(\text{AB})}_{i,j} + \mathcal{U}^{(\text{ATL})}_{i,j},
\end{equation*}
one last bivariate copula $C_5$ is integrated to represent the dependence between $\mathcal{U}^{(\text{AB+ATL})}_{i,j}$ and $\mathcal{U}^{(\text{ON})}_{i,j}$ which correspond to the Alberta-Atlantic cluster and the Ontario cluster. 
For a visual representation of the hierarchical copula used in the model, see Figure \ref{fig:imanconover}. 

This aggregation approach is consistent for instance with the work of \citet{Arbenz2012}. For a complete specification of the dependence model, their work includes a conditional independence assumption, meaning that given the aggregate scaled innovation at a given node, children of this node are independent from any node that is not a child of that given node. This same assumption holds in the current work which allows to fit any copula at each node regardless of the parametric family.

For more details about the copulas selected to compose the hierarchical copula model, refer to Section \ref{hierarchicalestim}.

Since losses from dependent business lines are not comonotonic, the total loss aggregated over all business lines is considered less risky than the set of all business line losses considered in silo (i.e. separately); this leads to the existence of a diversification benefit. The recognition of such diversification benefit for the risk adjustment for non-financial risks is allowed by IFRS 17 standards, when one is able to show that diversification holds in periods of stress.


\section{Implementation of the model}
\label{section:Imple}

The current section details the implementation 
of the proposed model. The estimation of the model parameters is first discussed. Then, we present a stochastic simulation algorithm to generate future loss ratios and obtain loss distributions in order to compute capital requirements and the risk adjustment for non-financial risks.


\subsection{Estimation algorithm}
\label{section:Estimation}

The current section details the estimation algorithms used for the estimation of the model parameters. The estimation is done in two steps. The first step consists in estimating parameters of the DGLM models representing the distribution of loss ratios independently for each business line. The second step entails specifying the structure of the hierarchical copula model and estimating its parameters.


\subsubsection{Marginal business line parameters estimation}

First, the parameters impacting a single business line are
estimated for each business line $k$ individually. The estimation approach relies on a Generalized Estimation Equations (GEE) method for parameters of the mean component of the DGLM and a Restricted Maximum Likelihood (REML) approach for the dispersion parameters.
The GEE is a convenient approach to estimate parameters of a DGLM model in the presence of correlation between observations, see for instance \citet{LongitudinalData,Hardin2013} and \citet{Smolarova2017}.

Indeed, the usual assumption of independence between observations of a DGLM does not hold in the current model. The REML allows circumventing a joint estimation of both mean and dispersion parameters and enables reducing the downward bias associated the traditional maximum likelihood estimates of dispersion parameters, see for instance \citet{REML}.

An iterative algorithm is used to obtain the estimates of the parameters. Indeed, parameters impacting the business line $k \in \{1,2,\ldots,K\}$ are split in two subsets: $\Theta^{(\mu)}_k$ and $\Theta^{(\phi)}_k$ which contain, respectively, the parameters driving the mean and dispersion:
\begin{eqnarray*}
\Theta^{(\mu)}_k &:=& \{ \iota^{(k)}, \alpha^{(k)}_{j}, \delta^{(k)}_{j}: j=2,\ldots,J\},
\\ \Theta^{(\phi)}_k &:=& \{ \iota_d^{(k)}, \gamma^{(k)}_{j} : j=2,\ldots,J\}.
\end{eqnarray*}
Note that the constraint $\alpha^{(k)}_{1}=\delta^{(k)}_{1}= \gamma^{(k)}_{1}=0$ is imposed to avoid identifiability issues.
The simultaneous estimation of the mean and dispersion for a Tweedie distribution is possible due to the statistical orthogonality of the parameters, see for instance \citet{Cox1987,Smyth1989}. For a fixed value of $k$, the iterative procedure goes as follows until convergence:

\begin{algorithm}[H]\label{algorithmdescent1}
\renewcommand{\thealgorithm}{Algorithm 1}
\vspace{0.21cm}
\caption{Estimates of the DGLM parameters}
\vspace{0.021cm}
\begin{algorithmic}\label{algorithmdescent}
\STATE{\textbf{Step (1)}} Keeping the current estimates of mean and dispersion parameters $\Theta^{(\mu)}_k$ and $\Theta^{(\phi)}_k$ fixed, refine the estimate of the development lag correlation parameter $\rho_k$.
\STATE{\textbf{Step (2)}} Keeping the current estimates of dispersion parameters $\Theta^{(\phi)}_k$ and development lag correlation parameter $\rho_k$ fixed, refine estimates of the mean parameters $\Theta^{(\mu)}_k$.
\STATE{\textbf{Step (3)}} Keeping the current estimates of mean parameters $\Theta^{(\mu)}_k$ and development lag correlation parameter $\rho_k$ fixed, refine estimates of the dispersion parameters $\Theta^{(\phi)}_k$.
\end{algorithmic}
\end{algorithm}

Details about the selection of a suitable value of the Tweedie index $p_k$ are provided in Appendix \ref{app:tweedieind}.

More details are now provided on each step of \ref{algorithmdescent}.
To ease the notation, loss ratios for a fixed accident semester $i \in \{1,2,\ldots,I\}$ are regrouped in a 
random vector: $\bm{Y}^{(k)}_{i}= \left[ Y^{(k)}_{i,1},\ldots,Y^{(k)}_{i,n_i} \right]^\top$, which corresponds to the vector of the $n_i:=J+1-i$ observed loss ratios from accident semester $i$ (i.e. for all development lags $j=1,...,n_i$). Recall that the unconditional distribution of each of its component is given by $Y^{(k)}_{i,j}\sim\text{TW}_{p_k}(\mu^{(k)}_{i,j},\phi^{(k)}_{j})$ for a development lag $j$. 

\vspace{\baselineskip}
\noindent \textbf{Step (1)}

At the first step, the correlation parameter is refined according to the following formula analogous to the sample correlation of scaled innovations:
\begin{equation*}
\label{eqn:autocorrelation}
\hat{\rho}_k=\cfrac{ \mathlarger{\sum}_{i=1}^{I-1} \mathlarger{\sum}_{j=2}^{n_i} \tilde{Y}^{(k)}_{i,j} \tilde{Y}^{(k)}_{i,j-1}}{\mathlarger{\sum}_{i=1}^{I-1} \mathlarger{\sum}_{j=2}^{n_i} \left(\tilde{Y}^{(k)}_{i,j-1} \right)^2 }.
\end{equation*}

\noindent \textbf{Step (2)}

The second step of \ref{algorithmdescent}, where mean parameters $\Theta^{(\mu)}_k$ are refined, is now discussed. Denote the column mean vector of $\bm{Y}^{(k)}_{i}$ by $\bm{\mu}^{(k)}_{i}= \left[ \mu^{(k)}_{i,1},\ldots,\mu^{(k)}_{i,n_i}\right]^\top$ and its dispersion vector $\bm{\phi}^{(k)}_{i}= \left[ \phi^{(k)}_{1},\ldots,\phi^{(k)}_{n_i} \right]^\top$. 
Mean parameters estimates that are being refined are chosen as the solution to the following GEE:
\begin{equation}
\label{eqn:GEE}
\mathlarger{\sum}_{i=1}^{n_i} D_i^{(k)^{\top}} \bm{V}_i^{(k)^{-1}} \left(\bm{Y}^{(k)}_{i}-\bm{\mu}^{(k)}_{i}\right)=0,
\end{equation}
where $D^{(k)}_i=\cfrac{\partial \bm{\mu}^{(k)}_i}{\partial \bm{\beta}^{(k)}}$ is 
a matrix of dimension $(n_i\times q)$ containing partial derivatives, $\bm{\beta}^{(k)}$ is the mean parameter vector of dimension $q = 1 + (I-1) + (J-1) = 2J-1$ given by
\begin{equation*}
    \bm{\beta}^{(k)}  = \begin{bmatrix}
\iota^{(k)} &
\alpha^{(k)}_{2} &
  \cdots&
  \alpha^{(k)}_{I}&
  \delta^{(k)}_{2}&
    \cdots&
     \delta^{(k)}_{J}
\end{bmatrix}_{(1 \times q)},
\end{equation*}
$\bm{V}^{(k)}_i= A^{(k)^{1/2}}_i R_{n_i}(\rho_k) A^{(k)^{1/2}}_i$ is a variance matrix of dimension $(n_i \times n_i)$, where $R_{n_i}(\rho_k) = R_{k,n_i}$ is the correlation matrix of the random vector $\bm{Y}^{(k)}_{i}$ defined in \eqref{eqn:corrmatrixfull}, and the diagonal matrix $A^{(k)}_i$ is given by
\begin{equation*}
\label{eqn:matrixA}
A^{(k)}_i=
\begin{bmatrix}
\phi^{(k)}_{1}V_k(\mu^{(k)}_{i,1}) & 0  & \ldots&0 \\
0&\phi^{(k)}_{2} V_k(\mu^{(k)}_{i,2}) &\ldots&0 \\
\vdots& \vdots & \ddots&\vdots\\
0& 0  &\ldots&\phi^{(k)}_{n_i}V_k(\mu^{(k)}_{i,n_i})\\
\end{bmatrix}_{(n_i \times n_i)},
\end{equation*}
with $V_k$ representing the variance function of the Tweedie family through equation $V_k(\mu)=\mu^{p_k}$. The variance matrix $\bm{V}^{(k)}_i$ is key to capture the correlation between observations since the correlation matrix $R_{n_i}(\rho_k)$ is functionally related to the scalar $\rho_k$. 

If the correlation matrix was the identity matrix, i.e., $R_{n_i}(\rho)=I_{n_i}$, where $I_{n_i}$ is the identity matrix of dimension $n_i$, the estimation procedure would be equivalent to the traditional DGLM estimation where the independence is assumed between observations. However, the independence assumption does not hold in the current study as outlined in Section \ref{section:DepWithin}.

The use of the Generalized Estimating Equation \eqref{eqn:GEE} is equivalent to using a weighted least squares estimator. This entails that the estimator of $\bm{\beta}^{(k)}$ is consistent. It is relevant to note that the combination of a DGLM model and a correlation structure between innovations is a novel addition to loss triangle modeling literature.

\vspace{\baselineskip}
\noindent \textbf{Step (3)}

The third step of the estimation procedure entails refining the estimate of variance dispersion parameters while keeping mean and correlation related parameter estimates fixed. For such purposes, an approach similar to \citet{Smyth1989} is followed, where the dispersion parameters estimation relies on the construction of an auxiliary GLM. In this auxiliary GLM, measures of disparity between realized and expected loss ratios, called deviances, are constructed and serve as the dependent variable. However, a modification to the latter approach proposed by \citet{REML} and implemented in the context of insurance claims modeling by \citet{Smythjorgensen2002} is considered. Such a modification entails applying a correction to the deviance associated with each observation based on its leverage; the correction allows reducing the downward bias of small sample dispersion parameter estimates, especially for development lags for which very few observations are available. \citet{REML} state that the leverage-based correction also provides the benefit of accelerating convergence of the estimation procedure while having an overall limited impact on resulting estimates.
The procedure is largely inspired by \citet{Smythjorgensen2002}, and the reader is referred to the latter paper for more extensive details. First, unit deviances are defined as
\begin{equation*}
\label{eqn:unitdev}
d^{(k)}_{i,j}
= 
2\left(y^{(k)}_{i,j}\cfrac{y_{i,j}^{(k)^{1-p_k}}-\mu_{i,j}^{(k)^{1-p_k}}}{1-p_k}-\cfrac{y_i^{(k)^{2-p_k}}-\mu_{i,j}^{(k)^{2-p_k}}}{2-p_k}\right).
\end{equation*}


The objective consists in choosing dispersion parameters such that the various $\phi^{(k)}_j$ defined in \eqref{eqn:dispmodel} match unit deviances $d^{(k)}_{i,j}$ as closely as possible; indeed, in the DGLM model, the expected value of the deviance $d^{(k)}_{i,j}$ is $\phi^{(k)}_j$ as stated in \citet{REML}.
The parameters and predictors of the auxiliary GLM constructed for dispersion parameter estimation are respectively given by the vector $\bm{\gamma}^{(k)}$ and the dummy matrix $Z$ defined according to
\begin{equation*}
\label{mat:cov.paramvec.disp}
Z\bm{\gamma}^{(k)}=
\begin{bmatrix}
	1&0&\hdots&0&0 \\
	1&1&\hdots&0&0 \\
	\vdots&\vdots&\vdots&\vdots&\vdots \\
	1&0&\hdots&0&1\\
	1&0&\hdots&0&0 \\
	\vdots&\vdots&\vdots&\vdots&\vdots \\
	1&0&\hdots&1&0 \\
	\vdots&\vdots&\vdots&\vdots&\vdots \\
	1&0&\hdots&0&0 \\
\end{bmatrix}_{(n \times J)}
\begin{bmatrix}
\iota_{d}^{(k)} \\
\gamma^{(k)}_{2}\\
\vdots\\
\gamma^{(k)}_{J}
\end{bmatrix}_{(J \times 1)}
=
\begin{bmatrix}
\iota_d^{(k)} \\
\iota_d^{(k)}+\gamma^{(k)}_{2} \\
\vdots\\
\iota_d^{(k)}+\gamma^{(k)}_{J}\\
\iota_d^{(k)}\\
\vdots\\
\iota_d^{(k)}+\gamma^{(k)}_{J-1}\\
\vdots\\
\iota_d^{(k)}
\end{bmatrix}_{(n \times 1)}.
\end{equation*}
where $n=\text{Card}( \mathcal{T}_U ) =\frac{J(J+1)}{2}$ is the number of observed elements in the loss triangle $\mathcal{T}_U$ associated with a given business line. 
Indeed, as seen in Figure \ref{fig:dglm}, each row of the matrix $Z$ corresponds to an entry of the loss triangle so that the element in the same row of $Z\bm{\gamma}^{(k)}$ contains the sum of all parameters characterizing its dispersion.

Moreover, the dummy matrix $X$ of dimension $n \times q$ is defined through
\begin{equation*}
\label{mat:cov.paramvec}
X\bm{\beta}^{(k)}=
\kbordermatrix{
&&&&I&&&&J&\\
&1&0&\hdots&0&0&0&\hdots &0 \\
&1&0&\hdots&0&1&0&\hdots &0 \\
&\vdots&\vdots&\vdots&\vdots&\vdots&\vdots&\vdots &\vdots \\
&1&0&\hdots&0&0&0&\hdots &1 \\
&1&1&\hdots&0&0&0&\hdots &0 \\
&1&1&\hdots&0&1&0&\hdots &0 \\
&\vdots&\vdots&\vdots&\vdots&\vdots&\vdots&\vdots &\vdots \\
&1&0&\hdots&1&0&0&\hdots &0 \\
}_{(n \times q)}
\begin{bmatrix}
\iota^{(k)} \\
\alpha^{(k–)}_{2} \\
  \vdots\\
  \alpha^{(k)}_{I}\\
  \delta^{(k)}_{2}\\
    \vdots\\
     \delta^{(k)}_{J}
\end{bmatrix}_{(q \times 1)}
=
\begin{bmatrix}
\iota^{(k)} \\
\iota^{(k)}+\delta^{(k)}_{2} \\
\vdots\\
\iota^{(k)}+\delta^{(k)}_{J} \\
\iota^{(k)}+\alpha^{(k)}_{2}\\
\iota^{(k)}+\alpha^{(k)}_{2}+\delta^{(k)}_{2} \\
\vdots\\
\iota^{(k)}+\alpha^{(k)}_{I}
\end{bmatrix}_{(n \times 1)},
\end{equation*}
where, as seen in Figure \ref{fig:glmeqn}, each row of the matrix in the right-hand side of the equation above corresponds to the sum of all coefficients characterizing the mean for a given entry of the loss triangle. Therefore, the matrix $X$ which contains predictors of the mean parameters GLM composing the DGLM model.
Moreover, another matrix $W$ of dimension $n\times n$ is defined through\footnote{Note that such definition of the weight matrix $W$ disregards the presence of correlation between observations across development lags. Indeed, the framework of \citet{Smythjorgensen2002} was developed under the assumption of independent observations. We leave the consideration of the correlation in this step as a future refinement to our model. }
\begin{equation*}
\label{eqn:weightmatrix}
W=\text{diag}\left[\left(\cfrac{\partial g(\mu^{(k)}_{i,j})}{\partial \mu}\right)^{-2} \cfrac{1}{\text{Var}(Y^{(k)}_{i,j})}\right]=\text{diag}\left(\frac{{\mu_{i,j}^{(k)}}^{2-p_k}}{\phi^{(k)}_j}\right),  \quad (i,j) \in \mathcal{T}_U
\end{equation*}
where diag is the operator putting elements of a sequence on the diagonal of a matrix.\footnote{The order of indices $(i,j)$ put into the diagonal of the matrix are respectively $(1,1)$, $(1,2), \ldots, (1,J), (2,1), (2,2), \ldots$}
This allows defining the diagonal projection matrix $H$ of dimension $n\times n$ as
\begin{equation*}
\label{eqn:projectionmat}
H=W^{1/2} X (X^{T} W X)^{-1} X^{T} W^{1/2}.
\end{equation*}
Elements on the diagonal of $H$, known as the leverages, are denoted by $h^{(k)}_{i,j}, (i,j) \in \mathcal{T}_U$.
The leverage matrix allows defining modified deviances as $d^{*(k)}_{i,j}=\cfrac{d^{(k)}_{i,j}}{1-h^{(k)}_{i,j}}$.
Using such modified deviances in the estimation procedures, the approach of \citet{Smyth1989} ultimately amounts to setting
\begin{equation*}
\bm{{\gamma}}^{(k)}=(Z^{T} W_d Z)^{-1} Z^{T} W_d \bm{z}_d
\end{equation*}
where $W_d$ is the $n \times n$ diagonal matrix defined as
\begin{equation*}
W_d=\text{diag}\left(\cfrac{1-h^{(k)}_{i,j}}{2}\right), \quad (i,j) \in \mathcal{T}_U
\end{equation*}
and the column vector $\bm{z}_d$ of length $n$ is given by
\begin{equation*}
{\bm{z}_d}= \left[\cfrac{d^{*(k)}_{i,j}-\phi^{(k)}_{j}}{\phi^{(k)}_{j}}+\log\phi^{(k)}_{j} \right]_{ (i,j) \in \mathcal{T}_U}.
\end{equation*}

Estimated parameters resulting from this three-step procedure can be found in Table found in Appendix \ref{app:parameters}.


\subsubsection{Hierarchical copula model specification and estimation}
\label{hierarchicalestim}

Once the marginal distribution parameters are estimated for all business lines, the estimation of the hierarchical model is then performed. To obtain the set of copula parameter estimates, maximum pseudo-likelihood is applied independently at each node of the copula hierarchical tree representation due to the conditional independence assumption of the hierarchical model mentioned in Section \ref{section:Hierarchical}. 

Under this method, model residuals are transformed as approximate Uniform$[0,1]$ variables, called the pseudo-uniform residuals, through the application of the decorrelated residual's empirical cdf to the decorrelated residual itself. 
This is equivalent to setting pseudo-uniform residuals equal to the scaled ranks decorrelated residuals.
More precisely, for a given accident year $i$ and development lag year $j$, pseudo uniform residuals $V$ and $\mathcal{V}$ are defined as
\begin{eqnarray*}
V^{(k)}_{i,j} &=& F_{k}\left( U^{(k)}_{i,j} \right), \quad k=1,\ldots,6,
\\ \mathcal{V}^{(ON)}_{i,j} &=& F_{(ON)}\left( \mathcal{U}^{(ON)}_{i,j} \right),  \quad \mathcal{V}^{(AB)}_{i,j} = F_{(AB)}\left(\mathcal{U}^{(AB)}_{i,j} \right)
\\ \mathcal{V}^{(ATL)}_{i,j} &=& F_{(ATL)}\left( \mathcal{U}^{(ATL)}_{i,j} \right),  \quad \mathcal{V}^{(AB+ATL)}_{i,j} = F_{(AB+ATL)}\left(\mathcal{U}^{(AB+ATL)}_{i,j} \right)
\end{eqnarray*}
where the $F$'s denote empirical cdfs (with a scaling correction to avoid values exactly equal to $1$), i.e.
\begin{equation*}
F_{k} (x) = \frac{  \sum_{(i,j) \in \mathcal{T}_U}  \mathds{1}_{ \{ U^{(k)}_{i,j} \leq x \} }   }{ \text{card}(\mathcal{T}_U) +1}.
\end{equation*}

Once the pseudo-uniform residuals are obtained, the parameters $\theta$ of a bivariate copula defining the dependence between to given groups of business lines $\kappa_1$ and $\kappa_2$ is estimated through
\begin{equation*}
\label{eqn:pseudolike}
\hat{\theta}= \underset{\theta}{\arg \max }\mathlarger{\sum}_{(i,j) \in \mathcal{T}_U} \log\left(c_{\theta}(v^{(\kappa_1)}_{i,j},v^{(\kappa_2)}_{i,j})\right),
\end{equation*} 
where $c_{\theta}$ is the parametric copula density and $v^{(\kappa)}_{i,j}$ denotes the pseudo-uniform residuals from the group of business lines $\kappa$.

The maximum pseudo-likelihood method differs from the traditional maximum likelihood estimate by not considering the parametric estimates of marginal distributions in the function to be maximized. Instead, one uses an empirical estimate of the marginal cumulative distribution functions. 

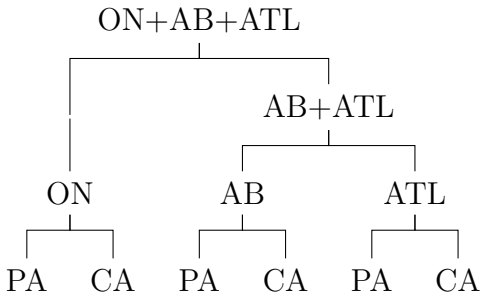
\begin{figure}[H]
		\centering
	\captionsetup{justification=centering}
	\begin{forest}
		forked edges,
		for tree={
			parent anchor=south,
			child anchor=north,
			if n children=0{
				tier=terminal,
			}{},
		}
	[ON+AB+ATL [,inner sep=0[ON [PA] [CA]]]
	[AB+ATL [AB [PA] [CA]]
	[ATL [PA] [CA]] ] ]
	\end{forest}
	\caption{HCM structure by province}
	\label{fig:hcm.final}
\end{figure}

\begin{figure}[H]
		\centering
\captionsetup{justification=centering}
\begin{forest}
	forked edges,
	for tree={
		parent anchor=south,
		child anchor=north,
		if n children=0{
			tier=terminal,
		}{},
	}
	[$\Pi$ [,inner sep=0[$\bm{t}_{8}$ [PA] [CA ]]]
	[$\bm{t}_{4}$ [$\bm{t}_5$ [PA] [CA] ]
	[$\Pi$ [PA] [CA]] ] ]
	\end{forest}
	\caption{HCM structure by copula family}
	\label{fig:hcm.final2}
\end{figure}
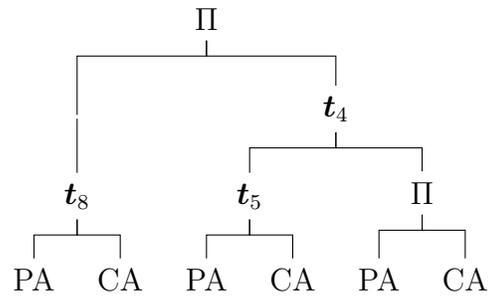

Figure \ref{fig:hcm.final} illustrates the hierarchy used during the aggregation process whereas Figure \ref{fig:hcm.final2} provides the bivariate copula family chosen at each step of the hierarchical aggregation. $\bm{t}_{\nu}$ represents the t-copula with $\nu$ degrees of freedom and $\Pi$ represents the independence copula. The bivariate t-copula with $\nu$ degrees of freedom and shape parameter $\rho$ is given by,
\begin{eqnarray*}
    C_{\nu,\rho}(u,v)&&=\bm{t}_{\nu,\rho}(t^{-1}_\nu(u),t^{-1}_\nu(v))\\[4pt]
    &&= \bigintssss_{-\infty}^{t^{-1}_\nu(u)} \bigintssss_{-\infty}^{t^{-1}_\nu(v)} \frac{1}{2\pi (1-\rho^2)^{1/2}}\left(1+\frac{s^2-2\rho st+t^2}{\nu(1-\rho^2)}\right)^{-\frac{\nu+2}{2}}dsdt,
\end{eqnarray*}
where $\bm{t}_\nu$ and $t_\nu$ are the multivariate and univariate distribution functions of a student-t, respectively.


\begin{multicols}{2}
\columnbreak
\end{multicols}


Table \ref{table:Copulagofprov} provides the parameter estimates for each bivariate copula in the hierarchical model. The results are obtained using the \texttt{copula} and \texttt{TwoCop} package in \texttt{R}.\footnote{The degrees of freedom $\nu$ is rounded to the nearest integer for the \textbf{copula} package.}
The column \textit{$p$-value} from the table provides the $p$-values from Cram\'{e}r-von Mises tests applied to verify the goodness-of-fit of the copulas, see \citet{Genest2008,Remillard2009}. The null hypothesis of the test is $H_0: C \in C_{\theta}$, i.e., the copula $C$ is indeed part of the parametric family $C_{\theta}$.
\vspace{1cm}
\begin{table}[H]
	\centering
	\begin{tabular}{ C{3cm}  C{2.5cm} C{3.5cm} C{3.5cm} C{2cm} }
		\toprule
		\centering{Province}&\centering{Copula family}&\centering{Dependence parameters}&Standard error of  $\rho$ &$p$-value\\
		\midrule
		ON&t&$\nu=8, \rho=0.166$&0.050&0.59\\
		AB&t&$\nu=5, \rho=0.290$&0.049&0.77\\
		ATL&Independence& - & - & 0.07 \\
		AB+ATL&t&$\nu=4, \rho=0.228$&0.050&0.39\\
		ON+AB+ATL&Independence& - & - & 0.54 \\
		\bottomrule
	\end{tabular}
	\caption{Parameters and goodness-of-fit of copula models by province or province group}
	\label{table:Copulagofprov}
\end{table}
\vspace{0.25cm}


\subsection{Simulation algorithm}
\label{section:Simulation}

The procedure to perform a stochastic simulation of unobserved loss ratios in the loss triangles based on the model outlined in Section \ref{section:Model} is provided in this section.

The first step consists in simulating independent decorrelated innovation vectors 
\begin{equation}
\label{uncorrInnovAllBL}
\mathbf{U}_{i,j} = \left[U^{(1)}_{i,j},\ldots,U^{(K)}_{i,j} \right]
\end{equation}
for all accident semester and development lag combinations $(i,j)$ that are yet unobserved, representing future observations.
 The dependence structure in the HCM is achieved through the Iman-Conover reordering algorithm proposed in \citet{Iman1982} and adapted by \citet{Arbenz2012}. Subsequently, the covariance structure of residuals across development lags is included in simulated innovations by applying a linear transformation,  which provides scaled innovations. The latter are finally transformed through an inversion procedure to obtain simulated values for all missing loss ratios.

A total of $N$ realizations
of the loss triangles must be performed. Each iteration consisting in the simulation of a single realization of the $K$ loss triangles involves the following steps:
\begin{enumerate}
	\item First, independently for each $(i,j) \in \mathcal{T}_L$, simulate the decorrelated innovation vector $\mathbf{U}_{i,j}$ defined in \eqref{uncorrInnovAllBL}, where the marginal distribution of each component $U^{(k)}_{i,j}$, $k=1,\ldots,K$ is standard normal and where the copula driving the dependence between elements of $\mathbf{U}_{i,j}$ is the aforementioned hierarchical copula model. Details on how to perform such a simulation are found in Appendix \ref{app:reorderexample}. Note that imposing the standard normal distribution to decorrelated innovations is an approximation justified by \eqref{theo:convergnorm}.
	\item The second step consists in inducing the correlation structure by applying a transformation on the decorrelated innovations so as to obtain the scaled innovations. This is done independently for each business line $k$ and accident semester $i$. Indeed, for each $k=1,\ldots,K$ and $i=2,\ldots,I$, define the vector
	$$\breve{\mathbf{U}}^{(k)}_{i} = \left[U^{(k)}_{i,I+2-i} , \ldots, U^{(k)}_{i,J}\right]^\top$$
	which contains the simulated decorrelated innovations associated to each unobserved loss ratio. Then, as explained in Appendix \ref{app:CondDistr}, the conditional distribution of unobserved loss ratio scaled innovations
$$\breve{\tilde{\mathbf{Y}}}^{(k)}_{i} = \left[\tilde{Y}^{(k)}_{i,I+2-i} , \ldots, \tilde{Y}^{(k)}_{i,J}\right]^\top$$ 
given observed loss ratios $\left[\tilde{Y}^{(k)}_{i,1} , \ldots, \tilde{Y}^{(k)}_{i,I+1-i}\right]^\top$ for the given accident year is approximately multivariate normal with mean vector $\breve{\mathbf{M}}^{(k)}_{i}$ and covariance matrix $\breve{V}^{(k)}_{i}$ as defined by \eqref{eq:condE}-\eqref{eq:condV} in Appendix \ref{app:CondDistr}.
This allows simulating the unobserved scaled innovations vector 
through
$$\breve{\tilde{\mathbf{Y}}}^{(k)}_{i} =  \breve{\mathbf{M}}^{(k)}_{i} + \breve{L}^{-1}_{k,i}\breve{\mathbf{U}}^{(k)}_{i}$$
where $\breve{L}^{-1}_{k,i}$ is the inverse of the lower triangle matrix in the Choleski decomposition of $\breve{V}^{(k)}_{i}$.
	\item Finally, the simulated scaled innovations are transformed such that they are properly scaled and that marginal distributions match the true Tweedie one from the model (instead of being Gaussian): 
\begin{equation*}
Y_{i,j}^{(k)}= F^{-1}_{i,j,k} \left( \Phi(\tilde{Y}^{(k)}_{i,j}) ;\mu^{(k)}_{i,j},\phi^{(k)}_{j},p^{(k)} \right) \approx TW_{p^{(k)}}(\mu^{(k)}_{i,j},\phi^{(k)}_{j})
\end{equation*}
where $F^{-1}_{i,j,k}$ is the functional inverse of the CDF of the marginal distribution of the loss ratio $Y_{i,j}^{(k)}$, and $\Phi$ is the standard normal CDF. Indeed, in the model, the unconditional distribution of the scaled innovations is approximately the standard normal one.
\end{enumerate}

Note that the correlation of loss ratios $Y^{(k)}_{i,j_1}$ and $Y^{(k)}_{i,j_2}$ in this simulation is only approximately equal to $\rho^{|j_1-j_2|}_k$; indeed, in the third step of the simulation algorithm above, when changing the marginal distribution of loss ratios from normal to Tweedie, the correlation structure also changes as the Pearson correlation between two random variables is not invariant to changes in their marginal distributions.

\section{Numerical Results}
\label{section:Numerical}

The current section illustrates how the stochastic model presented in Section~\ref{section:Model} can be used to calculate the insurer's risk adjustment for non-financial risks and its economic capital. Such calculation is performed through a Monte-Carlo simulation where multiple realizations of the unobserved (i.e. future) elements from the loss triangles are generated. Such realizations are used to construct a loss distribution for the insurer on which risk measures can be applied to quantify the insurer's exposure. This approach is referred to as the confidence level method. It is subsequently compared to the alternative CoC approach for the determination of the risk adjustment for non-financial risk. Capital requirement allocation approaches are also explored in the numerical results.

\subsection{Traditional risk measurement approaches}

The confidence level approach relies on the use of a risk measure to determine the risk adjustment amount. We recall the definition of two traditional risk measures commonly used in practice for such purpose, namely Value-at-Risk (VaR) and Tail Value-at-Risk (TVaR).

For a loss random variable $X$, its VaR at confidence level $\alpha$ is defined by
\begin{equation*}
\label{eqn:var}
\text{VaR}_{\alpha}(X)=\inf\{x\in\mathbb{R} \hspace{.1cm} | \hspace{.1cm} \mathbb{P}[X\leq x]\geq \alpha\}.
\end{equation*}
VaR$_{\alpha}(X)$ represents the $\alpha-$quantile of the loss distribution.
A drawback of $\text{VaR}$ is its blind spot for risk scenarios beyond the confidence level $\alpha$; this risk measure is not sufficient to understand the spectrum of worst possible losses for insurers. 

This points toward considering an alternative risk measure. TVaR at confidence level $\alpha$ is defined by
\begin{equation*}
\label{eqn:tvar}
\text{TVaR}_{\alpha}(X)=\cfrac{1}{1-\alpha}\bigintssss_\alpha^1 \mbox{VaR}_{u}(X)\text{d}u.
\end{equation*}
TVaR$_{\alpha}(X)$ can be interpreted as the expected loss over the worst $(1-\alpha)\%$ scenarios. 
This measure is meant to correct for the blind spot of VaR by providing additional insight on the behavior in the tail of the loss distribution. Indeed, TVaR takes into account potential outcomes beyond any chosen confidence level. Furthermore, TVaR is a coherent risk measure as it complies with properties established in \citet{Artzner1999}, see \citet{Acerbi} for the proof of coherence of TVaR.

Another way of determining the risk adjustment for non-financial risks is through the CoC method described in \citet{IAA2018}, where the risk adjustment is the present value of the future costs of capital associated with the unpaid claim liabilities. 
In the CoC approach, the risk adjustment for non-financial risks is calculated by
\begin{equation}
\label{eqn:costofcap}
\text{Risk adjustment}=\mathlarger{\sum}_{t=1}^{T} \cfrac{r_t \cdot C_t}{(1+d_t)^{t}}, 
\end{equation}
where $C_t$ represents the assigned capital amount for the period ending at time $t$, $r_t$ is the selected cost of capital rate for period ending at time $t$, $d_t$ is the selected discount rate allowing to discount from time $t$ back to time $0$, and $T$ is the number of periods considered. 

The main advantage of the cost of capital method is its simplicity and interpretability.
In the CoC method, the cost of bearing the uncertainty in the liabilities is reflected through the CoC rate, whereas it is represented by the loss distribution when using the confidence level approach. Nevertheless, when using the CoC approach, the IFRS 17 regulatory framework requires the risk adjustment to be converted to a VaR confidence level; the insurer is required to disclose such equivalent confidence level. Moreover, the CoC technique requires setting additional assumptions about the cost of capital rate.

A first simulation based on the loss triangle model and parameter estimates obtained in Sections \ref{section:Model} and \ref{section:Imple} is now performed to calculate the economic capital based on TVaR at level $\alpha=99\%$ of the aggregate loss distribution as recommended by the Canadian regulatory requirements, \citet{OSFI2019}. Note that \citet{OSFI2019} allows using either TVaR at level $\alpha=99\%$ or VaR at level $\alpha=99.5\%$ as the economic capital requirement. We focussed TVaR measures, as it is more representative of tail events, and it possesses more advantageous theoretical properties.
The aggregate discounted loss distribution obtained by generating $100,\!000$ scenarios of unpaid claim liabilities loss triangles is used to calculate capital requirements and its allocation to the six business lines. For illustrative purposes, in what follows, the discount and cost of capital rates are assumed to be constant for all periods; for all $t$, we have $d_t=2\%$ and $r_t=5\%$, respectively.

The results of this simulation are found in Tables \ref{table:ECAPsimul} and \ref{table:allriskmeasuresminusexp}.  
In the first row (\textit{Aggregate}) of Table~\ref{table:ECAPsimul}, the first six columns contain the allocation of the economic capital split to all business lines according to the Euler allocation principle, that is
\begin{equation*}
\label{eqn:TVaREuler}
\mbox{TVaR}_{\alpha}(X_i|S)=E(X_i|S>\mbox{VaR}_{\alpha}(S)),
\end{equation*}
where $S=X_1+...+X_6$. We refer the reader to \citet{Tasche1999} and \citet{McNeil2005} for detailed explanations concerning the TVaR-based allocation and the Euler allocation principle. Also, in this first row, the last column (\textit{Total}) provides the economic capital based on the TVaR at confidence level $\alpha=99\%$ applied to the empirical aggregate loss distribution, where the aggregate loss is obtained by summing discounted liabilities over all accident years, development lags and business lines. The second row (\textit{Silo}) of Table~\ref{table:ECAPsimul} provides the TVaR at confidence level $\alpha=99\%$ for each business line on a standalone basis. The element in the last column (\textit{Total}) for that second row is simply the sum of TVaRs over each business line.

A diversification benefit of \$482 million is therefore obtained by subtracting the Aggregate capital requirement from the Silo total capital requirement. Such an amount seems very modest due to being less than 0.5\% of the economic capital of the fictitious insurer. This is however explained in the current case by the fact that the exposure is highly concentrated in the Ontario Personal (ON/PA) insurance business line; diversification has a marginal impact due to the much lesser exposure to other lines of business. The diversification benefit would have been much higher if the insurer's exposure had been more balanced.


\begin{table}[H]
	\centering
	\begin{tabular}{L{1.5cm} c  c c   c c   c c | c}
		\toprule
		&&ON&ON&AB&AB&ATL&ATL&\multirow{2}{1.5cm}{Total}\\
		&&PA&CA&PA&CA&PA&CA&\\
		\midrule
		\multirow{2}{1cm}{TVaR$_{99\%}$}&Aggregate&  83,539 &  6,121 & 15,954  & 1,760 &  7,636 &   637 &115,647 \\
		&Silo &83,583 &  6,195&  16,169 &  1,811 &  7,712 &   659& 116,129 \\
		\bottomrule
	\end{tabular}
	\caption{Economic capital and allocation to business lines (millions CAD)}
	\label{table:ECAPsimul}
\end{table}

We now turn to the calculation of the risk adjustment for non-financial risks, which is summarized in Table~\ref{table:allriskmeasuresminusexp}. Two different methods are compared: the first based on TVaR at level $\alpha={87\%}$, and the second being the CoC method.

The first row (\textit{Aggregate}) of the TVaR panel presents the excess over the mean using the TVaR$_{87\%}$ of the total portfolio loss along with the allocation of that amount to the various business lines according to the Euler allocation principle. The confidence level chosen is $\alpha={87\%}$ because it leads to roughly similar results as the CoC method in terms of total portfolio risk adjustment for non-financial risks given the assumptions made. The second row (\textit{Silo}) of the TVaR panel presents the excess over the mean using the TVaR$_{87\%}$ of the loss distribution for each standalone business line, along with the sum of such values across all business lines in the column \textit{Total}. The CoC panel contains the risk adjustment for the total portfolio, along with the values for standalone business lines. For individual business lines, the capital $C_t$ considered is the standalone business line capital calculated from the simulations. The `Equivalent $\alpha$' row contains the confidence level for which the univariate (Silo) VaR would give the same amount than the CoC approach. In the CoC method \eqref{eqn:costofcap}, the capital requirement $C_t$ considered is $C_t=$VaR$_{99\%}(X_t)-E(X_t)$ where $X_t$ is the aggregate loss of year $t$. This is consistent with requirements of IFRS 17, see for instance p.59 \citet{IAA2018}.

Comparing the total portfolio risk adjustment for the Aggregate versus the Silo approach in Table~\ref{table:allriskmeasuresminusexp}, one observes that amounts obtained through the TVaR based approach are smaller. This is due to the diversification of risks. Moreover, one sees that for a comparable total capital amount obtained with the CoC and TVaR methods, the risk allocation across business lines exhibits less concentration with the CoC approach.

\begin{table}[H]
	\centering
	\begin{tabular}{L{4cm} c  c c   c c   c c | c}
		\toprule
		&&ON&ON&AB&AB&ATL&ATL&\multirow{2}{1cm}{Total}\\
		&&PA&CA&PA&CA&PA&CA&\\
		\midrule
		$E(X)$&  &82,502 & 6,109& 15,891 & 1,755  &7,629 &  637   \\
		\midrule
		\multirow{2}{*}{TVaR$_{87\%}(X) - E(X)$}& Aggregate & 617&   8  &41  & 3&   4 &  $<$1 &673  \\
		& Silo &643  &52 &166&  34&  50&  13 &958 \\
		\midrule
		CoC &&451& 45& 101& 23& 46& 10&676 \\ 
		\multicolumn{2}{l}{Equivalent $\alpha$ for VaR$_\alpha (X) - E(X)$} &87.79& 91.92 & 84.43& 86.80& 93.19& 89.35&  \\ 
		\bottomrule
	\end{tabular}
	\caption{Risk adjustments for non-financial risks and allocation to business lines (millions CAD)}
	\label{table:allriskmeasuresminusexp}
\end{table}

Table \ref{table:costofcapitalsensitivity} performs a sensitivity analysis on the risk adjustment for non-financial risks with respect to the cost of capital rate $r_t$. Outcomes stemming from the baseline value $r_t=5\%$ are compared to figures obtained with either $r_t=4\%$ or $r_t=6\%$. Such sensitivity is seen in the table to be quite material.

\begin{table}[H]
	\centering
	\begin{tabular}{ c  c c   c c   c c | c}
		\toprule
		Cost of capital rate&ON&ON&AB&AB&ATL&ATL&\multirow{2}{1cm}{Total}\\
		$r_t$&PA&CA&PA&CA&PA&CA&\\
        \midrule
        4\%&361& 36& 81& 18& 37&  8&541\\
        5\%&451& 45& 101& 23& 46& 10&676\\
        6\%&541& 53& 121& 27& 55& 12&809\\
		\bottomrule
	\end{tabular}
	\caption{Sensitivity of the risk adjustment for non-financial risks (millions CAD) to the cost of capital rate, \textit{ceteris paribus}}
	\label{table:costofcapitalsensitivity}
\end{table}


\section{Conclusion}
\label{section:Conclusions}

This article provides a statistical model for the prediction of loss ratios associated to liabilities for incurred claims risk of a multi-line property and casualty insurer. The model was designed based on a automibile insurance dataset from the General Insurance Statistical Agency for which a history of loss ratios was available for combinations of two business line types (i.e. personal versus commercial) and three geographical regions. The model possesses advantageous theoretical features allowing for the reproduction of empirical characteristics of loss ratios identified in the dataset. A Tweedie distributed Double Generalized Linear Model is used to represent the marginal distribution of loss ratios, where accident semester and development lag effects are taken into account when modeling both the mean and the dispersion of the distribution. An autocorrelation structure represents the loss ratio dependence across the various development lags for a given accident semester and business line, whereas the dependence across business lines is represented by a hierarchical copula model. 
A two-step estimation procedure is followed: parameters for each standalone business lines are estimated separately first through Generalized Estimating Equations, and then the hierarchical copula model is constructed based on previously obtained marginal business lines parameter estimates. 

The model developed herein serves many purposes and can be used for reserving (e.g. determination of the risk adjustment for non-financial risks), financial reporting and economic capital requirements calculations. A key attribute of our model is its consistency with IFRS 17 reporting standards; the dependence structure between loss ratios of the various business lines embedded in the model can be used to quantify the joint loss distribution across such business lines, hence allowing for the computation of the diversification benefit recognized under the IFRS 17 standards. The methodology for the quantification of the diversification benefit relies on a stochastic simulation using the loss ratio prediction model to generate multiple cash flow scenarios for the insurer. Risk measures can then be applied to the set of generated cash flow scenarios to measure either capital requirements or the risk adjustment for non-financial risks of the whole company and their allocation to the various business lines. 

The estimation procedure was applied on the current study's dataset to obtain parameter estimates. The latter served as inputs to a stochastic simulation experiment which illustrated the calculation of capital requirements and the risk adjustment for non-financial risks based on the TVaR risk measure. In this experiment, values obtained for the latter quantities were compared to a cost of capital approach. It was seen that the TVaR based method provided a capital allocation that exhibits more concentration to significant business lines, in comparison to the CoC method which spread the allocation more evenly across the lines.


\bibliography{bibliography}


\newpage
\appendix

\begin{appendices}


\section{Parameter estimates for marginal business lines}
\label{app:parameters}


Parameters for the accident semester and development lag effects which are denoted AS and DL, respectively. Furthermore, the lines of business are Personal Auto (PA) and Commercial Auto (CA) for the three regions of Ontario (ON), Alberta (AB) and Atlantic Canada (ATL). 


\begin{table}[H]
\begin{center}
\setlength\extrarowheight{-1pt}
\resizebox{14cm}{!}{
\begin{tabular}{c c c c c c c c}

\toprule

No.  &  Parameter  &   PA ON   &   CA ON   &   PA AB   &   CA AB   &   PA ATL   &   CA ATL   \\

\midrule

1 &  Intercept   & -1.55 & -1.80 & -1.05 & -1.12 & -1.40 & -1.48 \\
2 &  AS $=$ 2003-2  & -0.13 & 0.03 & 0.01 & -0.14 & -0.10 & -0.07 \\
3 &  AS $=$ 2004-1  & -0.29 & -0.24 & -0.21 & -0.36 & -0.27 & -0.29 \\
4 &  AS $=$ 2004-2  & -0.10 & -0.27 & -0.12 & -0.10 & -0.09 & -0.07 \\
5 &  AS $=$ 2005-1  & -0.23 & -0.40 & -0.21 & -0.42 & -0.10 & -0.48 \\
6 &  AS $=$ 2005-2  & -0.06 & 0.04 & -0.12 & -0.24 & 0.05 & -0.07 \\
7 &  AS $=$ 2006-1  & -0.11 & -0.23 & -0.26 & -0.37 & -0.17 & -0.34 \\
8 &  AS $=$ 2006-2  & 0.09 & 0.00 & -0.01 & -0.07 & 0.01 & -0.38 \\
9 &  AS $=$ 2007-1  & 0.02 & -0.16 & -0.31 & -0.42 & -0.19 & -0.54 \\
10 &  AS $=$ 2007-2  & 0.08 & 0.10 & -0.11 & -0.19 & -0.02 & -0.33 \\
11 &  AS $=$ 2008-1  & -0.02 & -0.07 & -0.25 & -0.45 & -0.23 & -0.47 \\
12 &  AS $=$ 2008-2  & 0.09 & 0.36 & -0.13 & -0.29 & -0.25 & -0.39 \\
13 &  AS $=$ 2009-1  & 0.06 & 0.18 & -0.38 & -0.84 & -0.21 & -0.59 \\
14 &  AS $=$ 2009-2  & 0.27 & 0.16 & -0.20 & -0.51 & 0.01 & -0.54 \\
15 &  AS $=$ 2010-1  & 0.16 & 0.03 & -0.53 & -0.61 & -0.10 & -0.57 \\
16 &  AS $=$ 2010-2  & 0.15 & 0.09 & -0.23 & -0.66 & 0.01 & -0.12 \\
17 &  AS $=$ 2011-1  & -0.08 & -0.12 & -0.44 & -0.60 & -0.18 & -0.64 \\
18 &  AS $=$ 2011-2  & 0.00 & 0.00 & -0.18 & -0.34 & 0.01 & -0.06 \\
19 &  AS $=$ 2012-1  & -0.13 & -0.06 & -0.29 & -0.63 & -0.16 & -0.73 \\
20 &  AS $=$ 2012-2  & -0.03 & 0.03 & -0.09 & -0.21 & 0.13 & -0.45 \\
21 &  AS $=$ 2013-1  & -0.13 & -0.14 & -0.26 & -0.24 & -0.15 & -0.36 \\
22 &  AS $=$ 2013-2  & 0.06 & 0.02 & -0.02 & -0.30 & 0.17 & 0.00 \\
23 &  AS $=$ 2014-1  & -0.10 & -0.06 & -0.25 & -0.63 & -0.09 & -0.07 \\
24 &  AS $=$ 2014-2  & 0.07 & 0.18 & 0.06 & -0.25 & 0.07 & -0.03 \\
25 &  AS $=$ 2015-1  & -0.03 & -0.09 & -0.13 & -0.48 & 0.05 & -0.17 \\
26 &  AS $=$ 2015-2  & 0.13 & 0.12 & 0.05 & -0.43 & 0.31 & -0.05 \\
27 &  AS $=$ 2016-1  & -0.02 & -0.09 & -0.18 & -0.66 & 0.09 & -0.31 \\
28 &  AS $=$ 2016-2  & 0.09 & 0.02 & 0.02 & -0.33 & 0.16 & -0.11 \\
29 &  AS $=$ 2017-1  & -0.12 & -0.15 & -0.27 & -0.47 & 0.01 & -0.12 \\
30 &  AS $=$ 2017-2  & 0.15 & 0.13 & -0.06 & -0.17 & 0.20 & 0.03 \\
\bottomrule
\end{tabular}}
\caption{Mean model - Accident semester effects}
\end{center}
\end{table}

\begin{table}[H]
\begin{center}
\setlength\extrarowheight{-1pt}
\resizebox{14cm}{!}{
\begin{tabular}{c c c c c c c c}
\toprule
No.  &  Parameter  &   PA ON   &   CA ON   &   PA AB   &   CA AB   &   PA ATL   &   CA ATL   \\
\midrule
31 &  DL $=$ 2  & -0.33 & -0.24 & -0.63 & -0.46 & -0.60 & -0.47 \\
32 &  DL $=$ 3  & -0.55 & -0.44 & -1.21 & -1.14 & -0.90 & -0.81 \\
33 &  DL $=$ 4  & -0.61 & -0.43 & -1.36 & -1.30 & -0.98 & -0.87 \\
34 &  DL $=$ 5  & -0.61 & -0.37 & -1.43 & -1.37 & -1.05 & -0.94 \\
35 &  DL $=$ 6  & -0.65 & -0.38 & -1.51 & -1.49 & -1.15 & -0.99 \\
36 &  DL $=$ 7  & -0.71 & -0.43 & -1.56 & -1.55 & -1.24 & -1.07 \\
37 &  DL $=$ 8  & -0.82 & -0.50 & -1.66 & -1.66 & -1.35 & -1.20 \\
38 &  DL $=$ 9  & -0.97 & -0.63 & -1.77 & -1.78 & -1.49 & -1.30 \\
39 &  DL $=$ 10  & -1.14 & -0.78 & -1.90 & -1.94 & -1.67 & -1.46 \\
40 &  DL $=$ 11  & -1.34 & -0.96 & -2.08 & -2.14 & -1.84 & -1.62 \\
41 &  DL $=$ 12  & -1.56 & -1.15 & -2.25 & -2.37 & -2.02 & -1.75 \\
42 &  DL $=$ 13  & -1.78 & -1.42 & -2.46 & -2.61 & -2.22 & -1.91 \\
43 &  DL $=$ 14  & -2.02 & -1.68 & -2.68 & -2.79 & -2.42 & -2.14 \\
44 &  DL $=$ 15  & -2.25 & -1.91 & -2.89 & -3.01 & -2.62 & -2.21 \\
45 &  DL $=$ 16  & -2.45 & -2.11 & -3.15 & -3.24 & -2.82 & -2.40 \\
46 &  DL $=$ 17  & -2.64 & -2.34 & -3.39 & -3.39 & -3.05 & -2.70 \\
47 &  DL $=$ 18  & -2.84 & -2.60 & -3.58 & -3.83 & -3.25 & -3.14 \\
48 &  DL $=$ 19  & -2.99 & -2.77 & -3.76 & -4.03 & -3.49 & -3.39 \\
49 &  DL $=$ 20  & -3.15 & -2.79 & -3.98 & -4.23 & -3.63 & -3.45 \\
50 &  DL $=$ 21  & -3.33 & -3.03 & -4.22 & -4.42 & -3.74 & -3.91 \\
51 &  DL $=$ 22  & -3.48 & -3.25 & -4.51 & -4.74 & -3.95 & -4.11 \\
52 &  DL $=$ 23  & -3.63 & -3.46 & -4.70 & -4.89 & -4.08 & -4.15 \\
53 &  DL $=$ 24  & -3.74 & -3.72 & -5.10 & -5.05 & -4.34 & -4.67 \\
54 &  DL $=$ 25  & -3.88 & -3.78 & -5.37 & -4.96 & -4.69 & -5.07 \\
55 &  DL $=$ 26  & -4.03 & -3.88 & -5.82 & -5.22 & -4.74 & -4.95 \\
56 &  DL $=$ 27  & -4.15 & -4.58 & -5.83 & -5.28 & -5.08 & -6.56 \\
57 &  DL $=$ 28  & -4.25 & -4.46 & -5.84 & -9.87 & -5.09 & -6.80 \\
58 &  DL $=$ 29  & -4.23 & -4.55 & -6.09 & -13.40 & -5.62 & -5.73 \\
59 &  DL $=$ 30  & -4.57 & -4.67 & -6.16 & -13.48 & -5.72 & -12.12 \\
\bottomrule
\end{tabular}}
\caption{Mean model - Development Lag effects}
\end{center}
\end{table}

\begin{table}[H]
\begin{center}
\setlength\extrarowheight{-1pt}
\resizebox{14cm}{!}{
\begin{tabular}{c c c c c c c c}
\toprule
No.  &  Parameter  &   PA ON   &   CA ON   &   PA AB   &   CA AB   &   PA ATL   &   CA ATL   \\
\midrule
60 &  Intercept   & -4.80 & -5.78 & -4.29 & -2.94 & -4.53 & -4.80 \\
61 &  DL $=$ 2  & 0.56 & -1.36 & -0.89 & -2.08 & -2.05 & -0.68 \\
62 &  DL $=$ 3  & 0.58 & -1.75 & -1.53 & -2.79 & -3.11 & -1.05 \\
63 &  DL $=$ 4  & -0.16 & -1.72 & -1.32 & -2.46 & -3.43 & -1.03 \\
64 &  DL $=$ 5  & -0.87 & -1.81 & -1.55 & -2.79 & -4.14 & -2.76 \\
65 &  DL $=$ 6  & -1.33 & -2.02 & -2.00 & -3.01 & -4.71 & -2.34 \\
66 &  DL $=$ 7  & -1.65 & -2.20 & -3.15 & -4.35 & -4.91 & -2.19 \\
67 &  DL $=$ 8  & -2.47 & -2.46 & -3.53 & -3.92 & -4.68 & -1.87 \\
68 &  DL $=$ 9  & -3.23 & -1.76 & -3.36 & -3.44 & -4.05 & -1.19 \\
69 &  DL $=$ 10  & -2.88 & -1.72 & -2.86 & -2.53 & -3.26 & -1.00 \\
70 &  DL $=$ 11  & -2.21 & -1.54 & -2.72 & -2.40 & -2.94 & -0.73 \\
71 &  DL $=$ 12  & -1.50 & -0.55 & -2.52 & -2.07 & -2.46 & -0.38 \\
72 &  DL $=$ 13  & -0.77 & -0.35 & -2.03 & -1.72 & -2.56 & -0.50 \\
73 &  DL $=$ 14  & -0.43 & -0.48 & -1.58 & -1.40 & -2.04 & 0.00 \\
74 &  DL $=$ 15  & -0.11 & -0.66 & -1.83 & -1.29 & -2.01 & 0.28 \\
75 &  DL $=$ 16  & 0.36 & -0.23 & -1.89 & -0.98 & -1.93 & -0.07 \\
76 &  DL $=$ 17  & 0.04 & 0.13 & -1.45 & -0.89 & -1.87 & -0.20 \\
77 &  DL $=$ 18  & -0.08 & 0.64 & -1.26 & -1.53 & -1.82 & -0.47 \\
78 &  DL $=$ 19  & 0.22 & 0.43 & -1.46 & -0.97 & -2.27 & -0.33 \\
79 &  DL $=$ 20  & -0.03 & 0.81 & -1.50 & -0.66 & -2.42 & 0.06 \\
80 &  DL $=$ 21  & 0.48 & 0.16 & -1.05 & -0.65 & -2.51 & 0.27 \\
81 &  DL $=$ 22  & 0.47 & 0.49 & -0.70 & -0.36 & -2.41 & 0.82 \\
82 &  DL $=$ 23  & 1.06 & 0.66 & -0.50 & -0.50 & -2.30 & 0.95 \\
83 &  DL $=$ 24  & 1.25 & 0.98 & -0.24 & -0.23 & -1.45 & 0.64 \\
84 &  DL $=$ 25  & 0.80 & 0.51 & -0.46 & -0.07 & -1.43 & 0.25 \\
85 &  DL $=$ 26  & 1.39 & 0.01 & -1.05 & 0.08 & -1.92 & 0.66 \\
86 &  DL $=$ 27  & 1.33 & -1.81 & -1.27 & 0.60 & -2.21 & -0.95 \\
87 &  DL $=$ 28  & 0.27 & -3.19 & -1.56 & -0.97 & -2.57 & -0.64 \\
88 &  DL $=$ 29  & 0.71 & -3.10 & -4.10 & -2.46 & -5.21 & 0.22 \\
89 &  DL $=$ 30  & 1.80 & -8.96 & -8.67 & 0.02 & -9.00 & -2.17 \\
\bottomrule
\end{tabular}}
\caption{Dispersion submodel - Development lag effects}
\end{center}
\end{table}

\begin{table}[H]
	\begin{center}
		\setlength\extrarowheight{-3.4pt}
		\resizebox{12.1cm}{!}{
			\begin{tabular}{ccccccc}
				\toprule
				Correlation &  PA ON  &  CA ON  &  PA AB  &  CA AB  &  PA ATL  &  CA ATL \\ 
				\midrule
				$\rho_k$ & 0.80 & 0.67 & 0.72 & 0.68 & 0.75 &  0.69\\ 
				\bottomrule
		\end{tabular}}
		\caption{Estimated correlation parameter $\rho_k$ for each business line $k$}
	\end{center}
\end{table}

\begin{table}[H]
	\begin{center}
		\setlength\extrarowheight{-3.4pt}
		\resizebox{12.1cm}{!}{
			\begin{tabular}{ccccccc}
				\toprule
				Index parameter &  PA ON  &  CA ON  &  PA AB  &  CA AB  &  PA ATL  &  CA ATL \\ 
				\midrule
				$p_k$ &  1.900  &  1.200  &  1.500  &  1.500 &  1.215  &  1.200\\ 
				\bottomrule
		\end{tabular}}
		\caption{Tweedie distribution index parameters $p_k$ for each business line $k$}
	\end{center}
\end{table}


\section{Selection of the Tweedie index $p_k$}
\label{app:tweedieind}


Estimating $p_k$ is not a trivial endeavour, and therefore a procedure inspired from \citet{DunnSmyth2004} is considered in the current work.

A set of fixed values of $p_k$, namely $p_k=\{1.105,1.110,1.115,\ldots,1.900\}$, is considered.
For each of these values, DGLM parameters $\Theta^{(\mu)}_k$ and $\Theta^{(\phi)}_k$ are estimated through maximum likelihood while assuming a null development lag correlation i.e. $\rho_k=0$, the latter assumption considerably simplifying the estimation. The value of $p_k$ for which the loglikelihood is maximized is the value selected as the parameter estimate.

Recall that values of $p_k$ must lie in the $(1,2)$ interval. However, for stability considerations, values below $1.1$ are not considered since values very close to $1$ tend to make the distribution multimodal; this complicates the estimation procedure and creates convergence issues. Moreover,
values close to $2$ were also disregarded for numerical considerations; when $p_k$ is close to $2$, the infinite sum approximation embedded in the Tweedie distribution includes a large number of terms that are materially different from zero, which makes computations more cumbersome.




\section{The Iman-Conover procedure}
\label{app:reorderexample}

Figure \ref{fig:imanconover} provides an illustration of the modeled dependence structure of the GISA dataset lines of business, which is based on a hierarchical copula. The Iman-Conover reordering algorithm is used to simulate from such copula in numerical experiments and it goes as follows:
\begin{enumerate}
	\item Simulate $k$ independent samples of size $m>>N$\footnote{In \citet{Cote2014} it is pointed out that the empirical distribution functions of the marginals and the copula converge asymptotically to the true distributions. Thus, a larger sample size $m$ provides a better estimate of the HCM sample.} composed of independent standard normal random variables: 
	\begin{equation*}
	 \mathbf{U}^{(k)} \sim N(0,1), \quad k=\{1,2,3,4,5,6\}.
	\end{equation*}
	\item Simulate independent copula samples of size $m$ from each bivariate copula $C_1,\ldots,C_5$.
\item Reorder the samples of each bivariate vector by merging the observed marginal ranks with the joint ranks in the copula sample. A brief example follows for the first node of the HCM.

\begin{table}[H]
	\centering
	\captionsetup{justification=centering}
\begin{tabular}{c c | c c | c c}
	\toprule
	$\mathbf{U}^{(1)}$ &Rank&$\mathbf{U}^{(2)}$&Rank&$C_1$&Ranks\\
	\midrule
	1.27&2&3.71&3&$(0.7,0.4)$&$(3,2)$\\
	-0.10&1&-2.19&1&$(0.2,0.9)$&$(1,3)$\\
	2.80&3&0.40&2&$(0.5,0.3)$&$(2,1)$\\
	\bottomrule
\end{tabular}
$\rightarrow$
\begin{tabular}{c}
	\toprule
	Reordered Sample\\
	\midrule
	$(2.80, 0.40)$\\
	$($-$0.10, 3.71)$\\
	$(1.27, $-$2.19)$\\
	\bottomrule
\end{tabular}
\caption{Iman-Conover reordering algorithm example for the first node of dependence structure (HCM) from Figure \ref{fig:imanconover}. Inspired by examples in \citet{Arbenz2012}}
\end{table}

Then, the reordered data is a sample from the copula $\left(\mathbf{U}^{(1)},\mathbf{U}^{(2)}\right)\sim C_1$.

\item Repeat step 3 for the first level copulas $C_2$ and $C_3$.
\item Aggregate the reordered data following the dependence structure to obtain samples from $\mathbf{U}^{(1)}+\mathbf{U}^{(2)}$ and respectively for $\mathbf{U}^{(3)}+\mathbf{U}^{(4)}$ and $\mathbf{U}^{(5)}+\mathbf{U}^{(6)}$.
\item Repeat step 3 to obtain sample from $\left(\mathbf{U}^{(3)}+\mathbf{U}^{(4)},\mathbf{U}^{(5)}+\mathbf{U}^{(6)}\right) \sim C_4$.
\item Aggregate the reordered sample from $C_4$ to obtain a sample from $\mathlarger{\sum}_{k=3}^{6}\mathbf{U}^{(k)}$, and repeat step 3 for $\left(\mathbf{U}^{(1)}+\mathbf{U}^{(2)},\mathlarger{\sum}_{k=3}^{6}\mathbf{U}^{(k)}\right) \sim C_5$.
\item To obtain a joint sample of $\left(\mathbf{U}^{(1)}, \mathbf{U}^{(2)}, \mathbf{U}^{(3)}, \mathbf{U}^{(4)}, \mathbf{U}^{(5)}, \mathbf{U}^{(6)}\right)$, perform the permutations applied to $\mathbf{U}^{(1)}+\mathbf{U}^{(2)}$ back to $\mathbf{U}^{(1)}$ and $\mathbf{U}^{(2)}$, the permutations applied to $\mathbf{U}^{(3)}+\mathbf{U}^{(4)}$ back to $\mathbf{U}^{(3)}$ and $\mathbf{U}^{(4)}$, and finally, the permutations applied to $\mathbf{U}^{(5)}+\mathbf{U}^{(6)}$ back to $\mathbf{U}^{(5)}$ and $\mathbf{U}^{(6)}$.
\item Get a subsample of size $N$ from the reordered sample of size $m$.
\end{enumerate}


\section{The conditional distribution of simulated scaled innovations}
\label{app:CondDistr}

The assumption made in the current paper's model based on \eqref{theo:convergnorm} is that for a given accident semester $i$ and business line $k$, the scaled innovations vector $\tilde{\mathbf{Y}}^{(k)}_i$ are approximately multivariate normal with a null mean vector and covariance matrix $R_{k,J}$ as defined in \eqref{eqn:corrmatrixfull}.

\vspace{\baselineskip}
A classic result on multivariate normal distributions is first recalled. Consider a multivariate normal random column vector $\textbf{X}$ which is decomposed into two blocks (i.e. two stacked random vectors): $\textbf{X} = [(\textbf{X}^{(1)})^\top \,\,  (\textbf{X}^{(2)})^\top]^\top$. Denote respectively the mean vector and covariance matrix of the entire vector $\textbf{X}$ and of each of the two blocks $\textbf{X}^{(1)}$ and $\textbf{X}^{(2)}$ by
\begin{equation*}
	\pmb{\mu} = \left[
				\begin{array}{c}
				\pmb{\mu}^{(1)}  \\
				\pmb{\mu}^{(2)}
				\end{array}
				\right], \quad 
	\Sigma = \left[
			\begin{array}{cc}
			\Sigma^{(1,1)} &  \Sigma^{(1,2)} \\
			\Sigma^{(2,1)} & \Sigma^{(2,2)}
			\end{array}
			\right].
	\end{equation*}
Then, the conditional distribution of $\textbf{X}^{(2)}$ given $\textbf{X}^{(1)}$ is multivariate normal with mean $\pmb{\mu}^{(2)} + \Sigma^{(2,1)} \left[\Sigma^{(1,1)}\right]^{-1} \left(\textbf{X}^{(1)}-\pmb{\mu}^{(1)}\right)$ and variance $\Sigma^{(2,2)} - \Sigma^{(2,1)} \left[\Sigma^{(1,1)}\right]^{-1} \Sigma^{(1,2)}$.

\vspace{\baselineskip}
We can decomposed the scaled innovation vector $\tilde{\mathbf{Y}}^{(k)}_i$ into two blocks: the unobserved one $\textbf{X}^{(1)} = \tilde{\mathbf{Y}}^{(k)}_{i,J+2-i:J} \equiv [Y^{(k)}_{i,J+2-i} , \ldots, Y^{(k)}_{i,J}]^\top$ and the observed one $\textbf{X}^{(2)} = \tilde{\mathbf{Y}}^{(k)}_{i,1:J+1-i} \equiv [Y^{(k)}_{i,1} , \ldots, Y^{(k)}_{i,J+1-i}]^\top$. In other words, 
\begin{equation*}
	\tilde{\mathbf{Y}}^{(k)}_i = \left[
				\begin{array}{c}
				\tilde{\mathbf{Y}}^{(k)}_{i,1:J+1-i}  \\
				\tilde{\mathbf{Y}}^{(k)}_{i,J+2-i:J}
				\end{array}
				\right].
	\end{equation*}
Since, the covariance matrix of $\tilde{\mathbf{Y}}^{(k)}_i$ can be decomposed as
\begin{equation*}
	R_{k,J} = \left[
			\begin{array}{cc}
			R_{k,J+1-i} &  R^{(1,2)}_{k,i} \\
			(R^{(1,2)}_{k,i})^\top & R_{k,i-1}
			\end{array}
			\right]
\end{equation*}
with 
\begin{equation*}
R^{(1,2)}_{k,i} \equiv 
\begin{bmatrix}
\rho^{J+1-i}_k & \rho^{J+2-i}_k & \rho^{J+3-i}_k & \ldots&\rho^{J-1}_k  \\
\vdots & \vdots & \vdots &\ddots&\vdots\\
\rho^{2}_k & \rho^{3}_k & \rho^{4}_k&\ldots&\rho^{i}_k  \\
 \rho_k & \rho^{2}_k & \rho^{3}_k&\ldots& \rho^{i-1}_k \\
\end{bmatrix}.
	\end{equation*}
	
Setting $\pmb{\mu} = \pmb{0}$ in the previous result along with
$\Sigma^{(1,1)} = R_{k,J+1-i}$, $\Sigma^{(2,2)} = R_{k,i-1}$ and $\Sigma^{(1,2)} = R^{(1,2)}_{k,i}$ leads to approximate the conditional distribution of unobserved scaled innovations $\tilde{\mathbf{Y}}^{(k)}_{i,J+2-i:J}$ given observed ones $\tilde{\mathbf{Y}}^{(k)}_{i,1:J+1-i}$ by a multivariate normal with mean vector and covariance matrix being respectively:
\begin{eqnarray}
    \breve{\mathbf{M}}^{(k)}_{i} \equiv \mathbb{E} \left[\tilde{\mathbf{Y}}^{(k)}_{i,J+2-i:J} \vert \tilde{\mathbf{Y}}^{(k)}_{i,1:J+1-i} \right] &=& (R^{(1,2)}_{k,i})^\top \left[ R_{k,J+1-i} \right]^{-1} \tilde{\mathbf{Y}}^{(k)}_{i,1:J+1-i}, \label{eq:condE}
    \\ \breve{V}^{(k)}_{i} \equiv \text{Cov} \left[\tilde{\mathbf{Y}}^{(k)}_{i,J+2-i:J} \vert \tilde{\mathbf{Y}}^{(k)}_{i,1:J+1-i} \right] &=& R_{k,i-1} -  (R^{(1,2)}_{k,i})^\top \left[ R_{k,J+1-i} \right]^{-1} R^{(1,2)}_{k,i} \!\!. \label{eq:condV}
\end{eqnarray}


\end{appendices}

\end{document}